\newif\ifTwoColumn
\newif\ifTechReport

\TechReporttrue    

\ifTwoColumn
	\documentclass[twocolumn]{autart}
	\usepackage{cite}
	\newcommand{\qedsymbol}{\hfill$\blacksquare$}
	\else
\ifTechReport
	\documentclass[eview,3p,number,11pt]{elsarticle_arXiv}
	\usepackage{amsthm}
	\usepackage{natbib}
	
	\usepackage{parskip}
	\usepackage{geometry}
	\renewcommand{\qedsymbol}{$\blacksquare$}
\else
	\documentclass[onecolumn]{autart}
	\usepackage{cite}
\fi
\fi

\usepackage{graphicx}          
\usepackage{amssymb}
\usepackage{amsmath}
\usepackage{bm}
\usepackage{stfloats}
\usepackage{graphicx}
\usepackage[ruled]{algorithm}
\usepackage{epstopdf}
\usepackage{textcomp}
\usepackage{subcaption}
\usepackage{enumerate}
\usepackage{subeqnarray}

\usepackage{xcolor,calc}

\newtheorem{myDef}{Definition}

\newtheorem{standing}{Standing Assumption}
\newtheorem{lemma}{Lemma}
\newtheorem{theorem}{Theorem}

\newtheorem{remark}{Remark}
\begin{document}

\begin{frontmatter}

\title{Identification of non-causal systems with random switching modes---EXTENDED VERSION} 


\author{Yanxin Zhang\textsuperscript{$\dagger$}, Chengpu Yu\textsuperscript{$\dagger$}, and Filippo Fabiani\textsuperscript{$\ddagger$}}

\address{$\dagger$~School of Automation, Beijing Institute of Technology, Beijing 100081, PR China}  
\address{$\ddagger$~IMT School for Advanced Studies Lucca, Piazza San Francesco 19, 55100, Lucca, Italy}                                           

\begin{abstract}                          
	We consider the identification of non-causal systems with random switching modes (NCS-RSM), a class of models essential for describing typical power load management and department store inventory dynamics. The simultaneous identification of causal-and-anticausal subsystems, along with the presence of random switching sequences, however, make the overall identification problem particularly challenging. To this end, we develop an expectation-maximization (EM) based system identification technique, where the E-step proposes a modified Kalman filter (KF) to estimate the states and switching sequences of causal-and-anticausal subsystems, while the M-step consists in a switching least-squares algorithm to estimate the parameters of individual subsystems. We establish the main convergence features of the proposed identification procedure, also providing bounds on the parameter estimation errors under mild conditions. Finally, the effectiveness of our identification method is validated through two numerical simulations.
\end{abstract}

\end{frontmatter}

\section{Introduction}
Non-causal switching dynamics arise in scenarios where actions depend on both historical and future states. In addition, these systems exhibit switching characteristics, potentially transitioning among different operational states, and thus leading to variations in the system behavior. In power systems, for instance, load management requires dedicated adjustments based on future demand \cite{Tan2023}. A controller can thus activate different modes to reduce load if a surge is anticipated, creating a dependency on future state. In traffic signal control \cite{Liao2024}, adaptive signal timing utilizes real-time vehicle flow predictions and synchronous historical/future data to enable autonomous phase adjustments without external centralized control. Its non-causal dependencies effectively prevent congestion propagation while enhancing traffic efficiency. In robotic systems \cite{Carloni2007}, collaborative robots may require anticipatory motion planning to avoid collisions, where future positions of other agents influence current decisions. In supply chain management, inventory provision decisions often depend on future demand and supplier lead times, creating non-causal dependencies between current actions and future states \cite{Liu2016}. Financial time series data often exhibit characteristics of sharp peaks and heavy tails. For instance, stock trading volumes may exhibit abnormal fluctuations prior to the release of significant announcements, which can be interpreted as the influence of future data on current observations \cite{El2023}. These systems exhibit switching behaviors due to discrete operational mode transitions (e.g., emergency shutdowns in power grids, traffic signal phase shifts, the anticipatory motion planning in collaborative robots), and their non-causality stems from the need to model feedback loops with delayed effects or predictive decision-making. Understanding and managing the complexity of these systems is therefore crucial for enhancing efficiency, reliability, and adaptability, enabling them to better meet the demands of industrial production and operations. This essentially motivates the interest in modeling, analyzing, and controlling such type of systems.


In several identification problems for dynamical systems, the input-output data are accompanied by temporal mode sequences. As the system's mode changes over time, each data point is associated with the active mode at that specific time. Hence, it is crucial to model the dynamics of different modes and infer transitions between modes \cite{Chan2008}. However, obtaining direct estimates of the dynamical system from input-output data is challenging and, in practice, prior knowledge about mode transitions is often unavailable. Therefore, estimating the switching behaviors poses a challenging yet highly significant problem that has attracted attention from researchers. Existing studies indeed propose algorithms to identify individual system dynamics and mode transition sequences from observed behaviors \cite{Ferrari2003}.

\subsection{Literature review}
Several works consider the identification of switching models \cite{Garulli2012,Bianchi2021}. Among various switching system models, jump Markov linear systems (JMLS) have emerged as a powerful framework for capturing abrupt random behavioral changes. These systems utilize a probabilistic structure in which mode transitions follow a Markov chain, effectively modeling stochastic switching dynamics through discrete state transitions with memoryless properties. In \cite{Mark2022}, a joint smoothing algorithm based on the expectation-maximization (EM) framework is proposed, with an E-step solution introduced to address exponential complexity in the JMLS. In \cite{Bemporad2018}, a numerically efficient two-step estimation method is developed, iteratively updating parameters and the switching sequence. The flexibility of this technique lies in its adaptability to various loss functions used in jump models, which significantly influence their shape and switching behavior.  Furthermore, the identification of jump Box–Jenkins (BJ) models is investigated in \cite{Piga2020}. These models consist of a finite collection of linear dynamical submodels in BJ form, switching over time according to a Markov chain. The system parameters are estimated iteratively using the Gauss-Newton and prediction error methods. In \cite{Sayedana2024}, a switching least-squares algorithm for autonomous Markov jump linear systems is proposed. Here, the authors provide a formal proof of the method’s strong consistency and establish its convergence rate as $\mathcal{O}(\sqrt{\log(T)/T})$, where $T$ is the time horizon. While existing literature primarily focuses on linear systems with Markov switching, these techniques may fail when mode transitions exhibit non-Markovian randomness. To address this limitation, methods for systems with random switching behavior have been proposed. For instance, \cite{Anna2018} employs a kernel-based approach to simultaneously solve estimation and classification problems in random switching systems. Similarly, \cite{Angelo2010} proposes a maximum-likelihood algorithm combining Kalman filtering and likelihood estimation to stabilize error convergence in general switched linear systems.

Furthermore, Gaussian mixture models (GMMs) served as a cornerstone for identifying switching systems \cite{Cao2016}, where EM-based algorithms are widely adopted to estimate latent modes and subsystem parameters \cite{Yang2012}. However, classical GMM frameworks are inherently limited to causal dynamics with single switching sequences, failing to address systems governed by bidirectional dependencies and dual independent switching behaviors. Our work extends GMM principles to non-causal switching dynamics, where outputs depend on both historical and future states through two distinct switching sequences. However, to the best of our knowledge, there is no literature on the system identification problem of non-causal systems with random switching modes (NCS-RSM). Although, there are some studies available on system identification for non-causal systems, such as the subspace \cite{Verhaegen1996} and the kernel methods \cite{Fang2024,Blanken2020}, these studies can only handle a single, non-causal system, rather than \textit{switching non-causal systems}. 

\subsection{Summary of contribution}

In this paper, we focus on the identification of NCS-RSM. The proposed method is developed under the expectation-maximization (EM) framework, which can be divided into two main parts. Specifically, in the E-step we adopt a Bayesian rule to compute the posterior estimate of the switching sequence, along with a modified Kalman filter (KF) for estimating the state of the causal and non-causal parts. In the M-step, instead, we propose a switching least-squares method to obtain the closed-form solution for the parameters and establish the convergence rate of the estimated parameters. Our main contributions can hence be summarized as follows:
\begin{enumerate}
	\item To the best of our knowledge, this is the first work considering the identification of NCS-RSM. To contrast with causal systems \cite{Bemporad2018,Piga2020}, we propose a modified Kalman filter (KF) with bidirectional estimation (forward for causal states, backward for non-causal ones) to resolve the non-causal dependency.
	\item Unlike prior works \cite{Mark2022,Sayedana2024}, which focus on causal systems with Markov switching, our method explicitly handles bidirectional switching dynamics (causal and non-causal subsystems) and random (non-Markov) switching sequences. Moreover, the switching sequences of the two directions is allowed to differ from each other. This enables modeling real-world scenarios where outputs depend on both past and future states.
	\item We show that our method has a $\mathcal{O}(\sqrt{\log(T)/T})$ rate of convergence under the assumptions of average stability and martingale difference noise. This result is generalized from causal Markov systems \cite{Sayedana2024} to non-causal random ones.
\end{enumerate}

\subsection{Paper organization}
The rest of the paper is organized as follows:
in Section \ref{sec3} we describe the considered system and formulate the related identification problem. In Section \ref{sec4}, instead, we discuss our EM method for the identification of the NCS-RSM, while in Section \ref{sec5} we provide its implementation details, as well as characterize the related convergence properties. Two simulation examples are finally discussed in Section \ref{sec6} to test the effectiveness of the proposed method numerically. 
The proofs of the technical results of the paper are all deferred to Appendix~\ref{sec:proofs}.

\textit{Notation:}
$\mathbb{Z}$ and $\mathbb{R}$ denote the set of integer and real numbers, respectively.
Given a matrix $X$, $\Vert X\Vert$ and $\Vert X\Vert_{\infty}$ denote respectively its spectral and infinity norms,  and $\textrm{tr}(X)$ denotes the trace. For a real symmetric matrix $P$, $\lambda_{\textrm{max}}(P)$ and $\lambda_{\textrm{min}}(P)$ are respectively its maximum and minimum eigenvalues. $\mathbb{P}[\cdot]$ and $\mathbb{E}[\cdot]$ respectively denote a probability distribution and the related expected value.  $\mathbb{P}_\theta[\cdot]$ denotes the probability density function with $\theta$ as parameters. $\mathbb{E}_\theta[\cdot]$ denotes the expectation operator with respect to the distribution $\mathbb{P}_\theta[\cdot]$. $\mathbb{S}^{n}$ is the space of $n \times n$ symmetric matrices and $\mathbb{S}_{\succ 0}^{n}$ ($\mathbb{S}_{\succcurlyeq 0}^{n}$) is the cone of positive (semi-)definite matrices. Given two square matrices $A$, $B$ of compatible dimension, $A\succcurlyeq B$ means that $A-B$ is positive semidefinite. For a random sequence $\{x(t)\}_{t\geq 1}$, $x(1:t)$ denotes $\{x(1),\cdots,x(t)\}$, $\sigma(x(1:t))$ denotes the sigma field generated by random variables $x(1:t)$. For a sequence $\{s_t\}_{t\in\mathbb{N}}$, $s_T=\mathcal{O}(T)$ indicates that $\lim\sup_{T\rightarrow\infty}s_T/T\textless\infty$, while $s_T=o(T)$ that $\lim\sup_{T\rightarrow\infty}s_T/T=0$. Finally, $I$ identifies a standard identity matrix. $\mathcal{N}(\mu,\Sigma^2)$ denotes the normal distribution of a random variable with mean $\mu$ and standard deviation $\Sigma$. In the remainder we will use Standing Assumption to postulate properties that hold throughout the paper.
 
\section{Mathematical formulation}\label{sec3}
We now describe the system concerned in this paper, together with the main assumptions, and successively formalize the problem to be addressed.
\subsection{System model description}
Consider the following discrete-time, non-causal system characterized by random switching modes:
\begin{subequations}\label{eq:1}
	\begin{align}
		&x_c(t)=A_c(s_c(t))x_c(t-1)+v_c(t),\\
		&x_a(t)=A_a(s_a(t))x_a(t+1)+v_a(t),\\
		&y(t)=C_c(s_c(t))x_c(t)+C_a(s_a(t))x_a(t)+v_m(t),
	\end{align}	
\end{subequations}
where $t\in\mathbb{Z}$ is the time instant, $x_c(t)\in\mathbb{R}^{n_{x_c}},x_a(t)\in\mathbb{R}^{n_{x_a}}$ are the causal and non-causal state vectors, respectively, $y(t)\in\mathbb{R}^{n_y}$ denotes the system output, while $s_c(t)\in\{1,2,\ldots,m_c\}\triangleq\Lambda_c$ and $s_a(t)\in\{1,2,\ldots,m_a\}\triangleq\Lambda_a$ are two discrete variables representing the possible switching modes. In addition, $v_c(t)\in\mathbb{R}^{n_{x_c}}$ and $v_a(t)\in\mathbb{R}^{n_{x_a}}$ are the system noise vectors, and $v_m(t)\in\mathbb{R}^{n_y}$ is the measurement noise vector. Finally, $A_c:\Lambda_c\to\mathbb{R}^{n_{x_c}\times n_{x_c}}$ and $A_a:\Lambda_a\to\mathbb{R}^{n_{x_a}\times n_{x_a}}$ denote the matrix functions associated to the causal and non-causal state dynamics, respectively, while $C_c:\Lambda_c\to\mathbb{R}^{n_{y}\times n_{x_c}}$ and $C_a:\Lambda_a\to\mathbb{R}^{n_{y}\times n_{x_a}}$ are those mapping the two state vectors to the measured output. Assume that the noise terms $v_c(t)$, $v_a(t)$ and $v_m(t)$ are distributed according to a Gaussian distribution with zero mean and finite variance $v_c(t)\sim\mathcal{N}(0,\Sigma_c(s_c(t))),
v_a(t)\sim\mathcal{N}(0,\Sigma_a(s_a(t))),
v_m(t)\sim\mathcal{N}(0,\Sigma_m).$

\begin{standing}\label{assumpt2}
	The NCS-RSM \eqref{eq:1} is stable in the average sense, which means
	\begin{align*}
		&\lim\sup_{T\rightarrow\infty}\frac{1}{T}\sum_{t=1}^{T}\Vert x_c(t)\Vert^2\textless \infty\quad\\
		&\lim\sup_{T\rightarrow\infty}\frac{1}{T}\sum_{t=1}^{T}\Vert x_a(T+1-t)\Vert^2\textless \infty
	\end{align*}
	where $T$ is the sample size of the available dataset.
\end{standing}
As a main consequence of Standing Assumption~\ref{assumpt2}, we note that the underlying stability requirement implies the existence of a stationary distribution for both causal and non-causal states, thereby ensuring the ergodicity of the system dynamics.

\begin{remark}
	Stability in the average sense is widely applied in linear systems \cite{Sayedana2024,sta_as1,sta_as2}. Compared to other commonly used notions (e.g., mean-square stability \cite{Long2023}, which requires $\lim_{T\rightarrow\infty}\mathbb{E}[\Vert x(T)\Vert^2]\textless 0$, and almost sure (a.s.) stability \cite{Cong2024}, ensuring $\mathbb{P}[\lim_{T\rightarrow\infty}\Vert x(T)\Vert=0]=1$), the assumption of stability in the average sense is weaker \cite{Sayedana2024}. Specifically, it only imposes convergence in a time-averaged sense, avoiding stricter pointwise or moment-based constraints.
\end{remark}
Assume the sample size is $T$. We denote the sigma field as $\mathcal{G}_t=\sigma(x_c(1:t))$ for the causal part,  $\mathcal{G}^{'}_{t}=\sigma(x_a(T:T+1-t))$ for the non-causal part, and $\mathcal{G}^{*}_t=\sigma(x_c(1:t),x_a(1:t))$ for the measured part.

\begin{standing}\label{assump1}
	The sequences of noise vectors $v_c(1:t), v_a(T:T+1-t), v_m(1:t)$ are the martingale difference sequences with respect to the filtrations $\{\mathcal{G}_t\}_{t\geq 1}, \{\mathcal{G}^{'}_t\}_{t\geq 1}, \{\mathcal{G}^{*}_t\}_{t\geq 1}$, respectively, i.e. $\mathbb{E}[v_c(t)\vert\mathcal{G}_{t-1}]=0$, $\mathbb{E}[v_a(t)\vert\mathcal{G}^{'}_{t-1}]=0$, $\mathbb{E}[v_c(t)\vert\mathcal{G}^{*}_{t-1}]=0$ and satisfy the following conditions:
	\begin{align*}
		&\lim\inf_{T\rightarrow\infty}\frac{1}{T}\sum_{t=1}^{T}v_k(t)v_k(t)^\top\succ 0,\quad k\in\{c,m\},\\
		&\lim\inf_{T\rightarrow\infty}\frac{1}{T}\sum_{t=1}^{T}v_a(T+1-t)v_a(T+1-t)^\top\succ 0.
	\end{align*}
\end{standing}

\begin{remark}
	A martingale is a sequence of random variables for which, at a particular time of the realized sequence, the expectation of the next value in the sequence is equal to the present observed value even given knowledge of all prior observed values.
	Standing Assumption \ref{assump1} denotes a common requirement for analyzing the convergence of system identification algorithms, enabling the noise process to exhibit non-stationary and heavy-tailed characteristics---see, e.g., \cite{Lai1982,PE2018,Chen1986}.
\end{remark}

The NCS-RSM in \eqref{eq:1} thus consists of two state equations and one output equation. Specifically, the first state equation represents the dynamics of the causal state variables, while the second one the dynamics of the non-causal state variables. The system output is determined by both the causal and non-causal states. Furthermore, both the causal and non-causal parts of the system are composed of multiple subsystems, and their corresponding switching sequences are different. Given some $T\in\mathbb{Z}$, which will denote the sample size of the available dataset, let the switching sequences of the causal and non-causal parts being denoted by $\bm{s}_c\triangleq\{s_c(t)\}_{t=1}^T$ and $\bm{s}_a\triangleq\{s_a(t)\}_{t=1}^T$, respectively. Each of them corresponds to a set of parameters, i.e., $s_c(t)=i$ determines the model parameter $\theta^c_i\triangleq\{A_c(i),C_c(i),\Sigma_c(i)\}$ that is active at the time instant $t$. In particular, the sequences $\bm{s}_c$ and $\bm{s}_a$ undergo random switches with certain (fixed) probabilities over time. Then, we have the following assumption:

\begin{standing}\label{assump3}
	The following conditions hold true:
	\begin{enumerate}
		\item The switching sequences $\bm{s}_c$, $\bm{s}_a$, and the subsystem parameters $\theta^c$, $\theta^a$ are all independent among them, i.e., $\mathbb P[\bm{s}_c\vert\theta^c]=\mathbb P[\bm{s}_c]$, $\mathbb P[\theta^c\vert \bm{s}_c]=\mathbb P[\theta^c]$,
		$\mathbb P[\bm{s}_a\vert\theta^a]=\mathbb P[\bm{s}_a]$, $\mathbb P[\theta^a\vert \bm{s}_a]=\mathbb P[\theta^a]$.
		\item The switching sequence follows a multinomial distribution, i.e., for any $t$, we have $\mathbb P[s_c(t)=i]=\pi^c_i, \quad i=1,\ldots,m_c$, $\mathbb P[s_a(t)=i]=\pi^a_i, \quad i=1,\ldots,m_a$,
		with $\sum_{i=1}^{m_c}\pi^c_i=1$, $\sum_{i=1}^{m_a}\pi^a_i=1$.
	\end{enumerate}
\end{standing}


The complete set of model parameters that comprehensively describe the NCS-RSM can be conveniently encapsulated into a parameter object $\theta$, defined as follows: 
\[
	\theta\triangleq\left\{ \{\theta^c_i\}_{i=1}^{m_c},\{\theta^a_i\}_{i=1}^{m_a},\{\pi^c_i\}_{i=1}^{m_c},\{\pi^a_i\}_{i=1}^{m_a},\Sigma_m\right\}.
\]

\subsection{Problem statement}
Our goal is hence to estimate the \textit{unknown} model parameters $\theta$ characterizing the NCS-RSM \eqref{eq:1} with the known state dimension, the initial states $x_c(0)$, $x_a(T+1)$, number of causal system modes $m_c$ and non-causal system modes $m_a$, together with a collection of noisy output measurements $\bm{y}$:
\[
	\bm{y}\triangleq\bm{y}_{1:T}=\{y(1),\ldots,y(T)\}.
\]

The NCS-RSM system contains both causal and non-causal components with distinct switching sequences. Addressing this problem faces two main challenges. First, the output depends on both causal and non-causal states, which are unmeasurable. Simultaneously identifying parameters for all subsystems is challenging due to their continuous switching patterns. Second, the system has two independent switching sequences. For example, at time $t$, the causal component might activate subsystem $i$ while the non-causal component activates subsystem $j$, creating $m_a\times m_c$ possible combinations.

In addition, the switching behavior of the subsystems is random and independent across different time instants, i.e., $\mathbb P[s_c(t)\vert s_c(t-1),\ldots,s_c(1)]=\mathbb P[s_c(t)]$, $\mathbb P[s_a(t)\vert s_a(t+1),\ldots,s_a(T)]=\mathbb P[s_a(t)], \quad t=1,\ldots,T$.
To deal with the identification problem of the NCS-RSM \eqref{eq:1}, the EM framework is adopted, which is an iterative method that can yield an estimate of the parameters at each iteration \cite{Dempster1977}. Let us denote the parameter estimate at the $k$-th iteration of the underlying algorithm as $\theta^k$. Then, the proposed method can be (qualitatively, for the moment) described by means of the following two steps:
\begin{enumerate}
	\item In the E-step, we develop a modified KF to estimate the states of the causal and non-causal parts. Furthermore, the Bayesian rule is used to obtain a posterior estimate of the switching sequence. Subsequently, the full-data likelihood function $Q(\theta,\theta^k)$ can be calculated.
	\item In the M-step, the likelihood function $Q(\theta,\theta^k)$ is maximized with respect to the parameters $\theta$. Then, the identification of the NCS-RSM is updated, yielding $\theta^{k+1}$.
\end{enumerate}

Next section will discuss in detail each step of the proposed technique for NCS-RSM identification.

\section{The EM method for identifying NCS-RSM}\label{sec4}
By making use of the dataset $\bm{y}$, we aim at estimating the system parameters $\theta$. To this end, a standard approach is to let coincide $\hat{\theta}$, i.e., our estimate of the true $\theta$, with a maximizer of the likelihood function, namely:
\begin{equation}\label{eq:mdf}
\hat{\theta}\in\underset{\theta}{\arg\max}~~\ln \mathbb P_\theta(\bm{y})~\text{ s.t. }~\eqref{eq:1},
\end{equation}
where we indicate with $\mathbb P_\theta(\bm{y})$ the probability density function of the output $\bm{y}$ given some sets of parameters $\theta$. Note that in switching systems, the likelihood function may exhibit multiple equivalent maxima due to the interchangeability of subsystem parameters (e.g., permuting subsystem labels can yield identical likelihood values). However, these equivalent solutions do not affect the system's physical properties. During model validation, we only need to compare whether the subsystem dynamic responses remain consistent across different permutations.

Let us denote the collection of state variables over $T$ as $\bm{x}_c\triangleq\{x_c(t)\}_{t=1}^T$ and $\bm{x}_a\triangleq\{x_a(t)\}_{t=1}^T$. Given any collection of data $\bm{y}$, note that the likelihood function $\ln \mathbb P_\theta(\bm{y})$, also called marginal density function of $\bm{y}$, can be decomposed into the following form:
\begin{align}\label{eq2.1}
		\ln \mathbb P_\theta(\bm{y})&=\ln \sum_{\bm{s_c}}\sum_{\bm{s_a}}\int\int\mathbb P_\theta[\bm{y}\vert\bm{x_a,x_c,s_a,s_c}]\nonumber\\
		&\hspace{3cm}\mathbb{P}_\theta[\bm{x_a,x_c,s_a,s_c}]d\bm{x_a}d\bm{x_c}\nonumber\\
		&=\ln \sum_{\bm{s_c}}\sum_{\bm{s_a}}\int\int\mathbb P_\theta[\bm{y}\vert\bm{x_a,x_c,s_a,s_c}]\nonumber\\
		&\hspace{2cm}\mathbb{P}_\theta[\bm{x_a,s_a}]\mathbb{P}_\theta[\bm{x_c,s_c}]d\bm{x_a}d\bm{x_c}
	\end{align}
	where the first equality follows from the law of total probability, which expands the marginal log-likelihood $\ln\mathbb{P}_\theta[\bm{y}]$ by marginalizing over all latent variables. The second equality stems from the mutual independence between the causal and non-causal subsystems.
	Since the states of both subsystems obey to Markovian dynamics and the outputs are conditionally independent given the switching mode, states and mode parameters, the posterior probability density function in \eqref{eq2.1} can be decomposed into the following structured forms:
\begin{align}\label{eq2.2}
	&\mathbb{P}_\theta[\bm{y}\vert\bm{x_a,x_c,s_a,s_c}]=\prod_{t=1}^{T}\mathbb{P}_\theta[y(t)\vert x_c(t),x_a(t),s_c(t),s_a(t)]\nonumber\\
	&\mathbb{P}_\theta[\bm{x}_c,\bm{s}_c]=\mathbb{P}_\theta[x_c(1),s_c(1)]\nonumber\\
	&\hspace{3cm}\prod_{t=2}^{T}\mathbb{P}_\theta[x_c(t)\vert x_c(t-1),s_c(t)]\mathbb{P}_\theta[s_c(t)]\nonumber\\
	&\mathbb{P}_\theta[\bm{x}_a,\bm{s}_a]=\mathbb{P}_\theta[x_a(T),s_a(T)]\nonumber\\
	&\hspace{1.7cm}\prod_{t=1}^{T-1}\mathbb{P}_\theta[x_a(t)\vert x_a(t+1),s_a(t)]\mathbb{P}_\theta[s_a(t)].
\end{align}
In the NCS-RSM \eqref{eq:1}, the state variables $\bm{x}_c,\bm{x}_a$ are governed by the switching sequences $\bm{s}_c,\bm{s}_a$. Direct maximization of the marginal log-likelihood $\ln \mathbb{P}_\theta[\bm{y}]$ is inherently challenging due to its nonconvexity and high-dimensional nature. Furthermore, as shown in the decomposition \eqref{eq2.1}, evaluating  $\mathbb{P}_\theta[\bm{y}]$ requires summation over all possible realizations of $\bm{s}_c,\bm{s}_a$, which exponentially increases the computational complexity of solving \eqref{eq:mdf}.

Another way to marginalize the latent variables (such as $\bm{x}_c, \bm{x}_a, \bm{s}_c, \bm{s}_a$) is by taking the expectation over these latter. Instead of maximizing the incomplete likelihood function $\ln \mathbb P_\theta(\bm{y})$, we can estimate the conditional density of the hidden variables given the observations $\bm{y}$ and an estimate of parameter $\hat{\theta}$. Then, parameter estimate $\hat{\theta}$ can be obtained by maximizing the complete likelihood function. The full-data complete likelihood function can be expressed as follows:
\begin{align}\label{eq:1.3}
	\ln \mathbb P_{\theta}[\bm{y},\bm{x}_c,\bm{s}_c,\bm{x}_a,\bm{s}_a]= &\ln \mathbb P_\theta[\bm{y}]\nonumber\\
	&+\ln \mathbb P_\theta[\bm{x}_c,\bm{s}_c,\bm{x}_a,\bm{s}_a\vert\bm{y}].	
\end{align}
This relation directly links $\mathbb P_\theta(\bm{y})$ and $\mathbb P_\theta[\bm{y},\bm{x}_c,\bm{s}_c,\bm{x}_a,\bm{s}_a]$, with the latter depending on the unknown states $\bm{x}_c$, $\bm{x}_a$ and switching sequences $\bm{s}_c$, $\bm{s}_a$. The key step is then to approximate $\ln \mathbb P_\theta[\bm{y}]$ by the above relation \eqref{eq:1.3}, where $\bm{x}_c$, $\bm{s}_c$, $\bm{x}_a$, and $\bm{s}_a$ can be approximated by their conditional expectations based on the observed data $\bm{y}$. 
Therefore, at each iteration $k$ of our EM-based algorithm, given the estimate $\theta^k$, the conditional expectation of $\ln \mathbb P_{\theta^k}[\bm{x}_c,\bm{s}_c,\bm{x}_a,\bm{s}_a\vert\bm{y}]$ is abbreviated as:
\begin{align}\label{eq2.4}
	\mathbb{E}_{\theta_k}[\cdot]=\int\int\sum_{\bm{s_c}}\sum_{\bm{s_a}}(\cdot)\mathbb P_{\theta^k}[\bm{x}_c,\bm{s}_c,\bm{x}_a,\bm{s}_a\vert\bm{y}]d(\bm{x}_c)d(\bm{x}_a).
\end{align}
Then, by applying the expectation operator $\mathbb{E}_{\theta_k}[\cdot]$ to both sides of \eqref{eq:1.3}, one obtains:
\begin{align*}
	&\mathbb{E}_{\theta^k}[\ln \mathbb P_\theta[\bm{y},\bm{x}_c,\bm{s}_c,\bm{x}_a,\bm{s}_a]]\\
	&\qquad\qquad=\mathbb{E}_{\theta^k}[\ln \mathbb P_\theta(\bm{y})]+\mathbb{E}_{\theta^k}[\ln \mathbb P_\theta[\bm{x}_c,\bm{s}_c,\bm{x}_a,\bm{s}_a\vert\bm{y}]]\\
	&\qquad\qquad=\ln \mathbb P_\theta(\bm{y})+\mathbb{E}_{\theta^k}[\ln \mathbb P_\theta[\bm{x}_c,\bm{s}_c,\bm{x}_a,\bm{s}_a\vert\bm{y}]].
\end{align*}
Define $Q(\theta,\theta^k)=\mathbb{E}_{\theta^k}[{\ln \mathbb P_\theta[\bm{y},\bm{x}_c,\bm{s}_c,\bm{x}_a,\bm{s}_a]}]$,
$V(\theta,\theta^k)=\mathbb{E}_{\theta^k}[\ln \mathbb P_\theta[\bm{x}_c,\bm{s}_c,\bm{x}_a,\bm{s}_a\vert\bm{y}]].$
The EM approach iteratively estimates the parameters in the following two steps. First, we compute the expectation $Q(\theta,\theta^k)$ based on $\theta^k$ obtained from the previous iteration. Under Standing Assumption \ref{assump3}, the full-data complete likelihood function can be decomposed by using the Bayesian rule, which is given as follows:
\begin{align}\label{eq2.5}
		&\ln \mathbb P_\theta[\bm{y},\bm{x}_c,\bm{s}_c,\bm{x}_a,\bm{s}_a]=\ln \mathbb{P}_\theta[\bm{y}\vert \bm{x}_c,\bm{s}_c,\bm{x}_a,\bm{s}_a]\nonumber\\
		&\hspace{2cm}+\ln \mathbb{P}_\theta[\bm{x}_c,\bm{s}_c]+\ln\mathbb{P}_\theta[\bm{x}_a,\bm{s}_a]	
	\end{align}
where the posterior probabilities are shown in \eqref{eq2.2}.
In view of the white noise assumption characterizing the disturbance affecting both state variables and measured output, note that the distribution of the these variables, given the subsystem modes $s_c(t)=j$, $s_a(t)=l$, is Gaussian too and given as follows:

\begin{align}\label{eq2.6}
	\mathbb P_\theta[y(t)\vert &x_c(t),x_a(t),s_c(t)=j,s_a(t)=l]=\vert 2\pi\Sigma_m\vert^{-1/2}\nonumber\\
	&\exp\{(y(t)-\mu_1(t,j,l))^\top\Sigma_m^{-1}(y(t)-\mu_1(t,j,l))\},\nonumber\\
	\mathbb P_\theta[x_c(t)\vert &x_c(t-1),s_c(t)=j]=\vert 2\pi\Sigma_c(t)\vert^{-1/2}\nonumber\\
	&\exp\{(x_c(t)-\mu_2(t,j))^\top\Sigma_c^{-1}(t)(x_c(t)-\mu_2(t,j))\},\nonumber\\
	\mathbb P_\theta[x_a(t)\vert &x_a(t+1),s_a(t)=l]=\vert 2\pi\Sigma_a(t)\vert^{-1/2}\nonumber\\
	&\exp\{(x_a(t)-\mu_3(t,l))^\top\Sigma_a^{-1}(t)(x_a(t)-\mu_3(t,l))\},
\end{align}
where $\mu_1(t,j,l)=C_c(j)x_c(t)+C_a(l)x_a(t),
	\mu_2(t,j)=A_c(j)x_c(t-1),
	\mu_3(t,l)=A_a(l)x_a(t+1).$
Denote $w_{tj}^c$ as the posterior probability of the switching sequence given the $\theta^k$, dataset $\bm{y}$ and $s_c(t)=j$ ($w_{tl}^a$ is defined similarly)
\begin{align}
	w^c_{tj}= \mathbb P_{\theta^k}[s_c(t)=j\vert\bm{y}]=\frac{\mathbb P_{\theta^k}[\bm{y}\vert s_c(t)=j]\pi^c_j}{\sum_{t=1}^{T} \mathbb P_{\theta^k}[\bm{y}\vert s_c(t)=j]\pi^c_j},\nonumber\\
	w^a_{tl}= \mathbb P_{\theta^k}[s_a(t)=l\vert\bm{y}]=\frac{\mathbb P_{\theta^k}[\bm{y}\vert s_a(t)=l]\pi^a_l}{\sum_{t=1}^{T} \mathbb P_{\theta^k}[\bm{y}\vert s_a(t)=l]\pi^c_l}.\label{eq:posterior_a}
\end{align}
Then, the objective function $Q(\theta,\theta^k)$ can be computed by using together \eqref{eq2.2}, \eqref{eq2.4}, \eqref{eq2.5}, \eqref{eq2.6}, and \eqref{eq:posterior_a}:
\begin{align}\label{eq:1.4}
	\mathbb{E}_{\theta^k}[{\ln \mathbb P_\theta[\bm{y},\bm{x}_c,\bm{s}_c,\bm{x}_a,\bm{s}_a]}]
	=\sum_{i=1}^3 Q_i(\theta,\theta^k),
\end{align}
where  
\begin{align}\label{eq2.3}
	&Q_1(\theta,\theta_k)=\mathbb{E}_{\theta^k}[\ln\mathbb{P}_\theta[\bm{y}\vert\bm{x}_c,\bm{s}_c,\bm{x}_a,\bm{s}_a]]\nonumber\\
	&=\sum_{t=1}^{T}\sum_{j=1}^{m_c}\sum_{l=1}^{m_a}w_{tj}^c w_{tl}^a\mathbb{E}_{\theta^k}[\ln\mathcal{N}(\mu_1(t,j,l),\Sigma_m)]\nonumber\\
	&Q_2(\theta,\theta^k)=\mathbb{E}_{\theta^k}[\ln\mathbb{P}_\theta[\bm{x}_c,\bm{s}_c]]\nonumber\\
	&=\sum_{t=1}^{T}\sum_{j=1}^{m_c}w_{tj}^c\mathbb{E}_{\theta^k}[\ln\mathcal{N}(\mu_2(t,j),\Sigma_c(j))+\sum_{j=1}^{m_c}\ln\pi_j^c\sum_{t=1}^{T}w_{tj}^c\nonumber\\
	&Q_3(\theta,\theta^k)=\mathbb{E}_{\theta^k}[\ln\mathbb{P}_\theta[\bm{x}_a,\bm{s}_a]]\nonumber\\
	&=\sum_{t=1}^{T}\sum_{l=1}^{m_a}w_{tl}^a\mathbb{E}_{\theta^k}[\ln\mathcal{N}(\mu_3(t,l),\Sigma_a(l))+\sum_{l=1}^{m_a}\ln\pi_l^a\sum_{t=1}^{T}w_{tl}^a.
	\end{align}

Subsequently, the second step is to maximize the $Q(\theta,\theta^k)$ to obtain $\theta^{k+1}$, formally defined as $\theta^{k+1}=\arg\max_\theta Q(\theta,\theta^k)$.

Algorithm~\ref{alg:EM} summarizes the two main steps of the proposed identification methodology for NCS-RSM. We characterize next the monotonic properties of the likelihood function in \eqref{eq:mdf} when the EM algorithm is iteratively applied to estimate the system parameters $\theta$:
\begin{lemma}\label{lemma:1}
Given a dataset $\bm{y}$, let $\{\theta^k\}_{k\in\mathbb{Z}}$ be the sequence generated by Algorithm~\ref{alg:EM}. Then, the likelihood function in \eqref{eq:mdf}, evaluated along $\{\theta^k\}_{k\in\mathbb{Z}}$, is non-decreasing, thereby yielding $\ln \mathbb P_{\theta^{k+1}}[\bm{y}]\geq\ln \mathbb P_{\theta^k}[\bm{y}]$ for all $k\in\mathbb{Z}$.
\end{lemma}
The proof of Lemma \ref{lemma:1} is shown in \ref{sec:proofs}.
\begin{algorithm}[h!]
	\caption{EM-based identification of NCS-RSM}\label{alg:EM}
	\smallskip
	
	\textbf{Initialization:} Collect data $\bm y_{1:T}$, set $\theta^0$
	
	\smallskip
	
	\textbf{Iteration} $k\in\mathbb{Z}$\textbf{:}
	\smallskip
	\begin{enumerate}
		\item \textbf{E-step}: Compute $Q(\theta,\theta^k)$ using \eqref{eq:posterior_a}, \eqref{eq:1.4}, \eqref{eq2.3}
		\smallskip
		\item \textbf{M-step}: Set $\theta^{k+1}=\underset{\theta}{\arg\max}~Q(\theta,\theta^k)$
	\end{enumerate}
\end{algorithm}

\section{Implementation details of the EM algorithm}\label{sec5}
We now delve into the details of the steps outlined in Algorithm~\ref{alg:EM}, ultimately establishing our main technical result characterizing the sample complexity of the proposed identification technique for NCS-RSM.

\subsection{The E-step}\label{subsec:E-step}
This step requires the calculation of the objective function $Q(\theta,\theta^k)$. Specifically, this shall be achieved on the basis of the parameter $\theta^k$ estimated in the previous iteration. Then, according to the expression of $Q(\theta,\theta^k)$ in \eqref{eq:1.4}, the expectations of states $\bm{x}_c,\bm{x}_a$ and the switching sequences $\bm{s}_c,\bm{s}_a$ given the data $\bm{y}$ are required. 

First, we calculate the posterior estimates of the switching sequences $\bm{s}_c$ and $\bm{s}_a$ by leveraging the Bayesian rule, namely $\mathbb P_\theta[s_c(t)\vert y(t)]=\mathbb P_\theta[s_c(t),y(t)]/\mathbb P_\theta[y(t)]$ and $\mathbb P_\theta[s_a(t)\vert y(t)]=\mathbb P_\theta[s_a(t),y(t)]/\mathbb P_\theta[y(t)]$. In addition, according to the formula of total probability one obtains $\mathbb P_\theta[y(t)]=\sum_{j=1}^{m_c} \mathbb P_\theta[y(t)\vert s_c(t)=j]\pi_j^c$, $\mathbb P_\theta[y(t)]=\sum_{l=1}^{m_a} \mathbb P_\theta[y(t)\vert s_a(t)=l]\pi_l^a$.

Then, the data point can be assigned to each subsystem at time $i$ by solving the following optimization problem:
\begin{align*}
	\hat{s}_c(t)&=\underset{j\in\{1,\ldots,m_c\}}{\arg\max}~\mathbb P_\theta[y(t)\vert s_c(t)=j]\pi_j^c,\\
	\hat{s}_a(t)&=\underset{l\in\{1,\ldots,m_a\}}{\arg\max}~\mathbb P_\theta[y(t)\vert s_a(t)=l]\pi_l^a,
\end{align*}
where maximizing $\mathbb P_\theta[y(t)\vert s_c(t)=j]\pi_j^c$ is equivalent to maximizing the posterior probability of $\mathbb P_\theta[s_c(t)=j\vert y(t)]$ which is commonly used for data classification.
In this work, we make use of a hard assignment variant of the EM algorithm as in \cite{Andrea2020}. This approach approximates the posterior distribution of the switching sequence by its maximum a posteriori (MAP) estimate, effectively collapsing the posterior probability to a Dirac delta function. Such a simplification is justified under the assumption that the posterior distribution $\mathbb{P}_\theta[s_c(t)=j\vert \bm{y}]$ is sharply peaked around the MAP estimate $\hat{s}_c(t)$, which often holds when switching probabilities are highly concentrated (the same holds for the non-causal part). Similar approaches have been adopted in \cite{Pal2024} for switching systems, where hard assignments reduce computational complexity while preserving estimation accuracy. Following the hard EM framework, the posterior probabilities $w_{ij}^c,w_{il}^a$ are approximated by their MAP estimates, resulting in a binary assignment for any $(t,j)\in\{1,\ldots,T\}\times\{1,\ldots,m_c\}$ (or $(t,l)\in\{1,\ldots,T\}\times\{1,\ldots,m_a\}$):
\[
	w_{tj}^c= \left\{
	\begin{array}{ll}
		1 & \text{if $\hat{s}_c(t)=j$} \\
		0 & \text{else}
	\end{array}
	\right.,\quad
	w_{tl}^a= \left\{
	\begin{array}{ll}
		1 & \text{if $\hat{s}_a(t)=l$} \\
		0 & \text{else}
	\end{array}
	\right.
\]

Successively, we focus on the reconstruction of the state variables $\bm{x}_c$ and $\bm{x}_a$, a task that is traditionally accomplished by means of a Kalman filter.  
Adapting the KF to our problem, however, requires few key modifications due to the dynamics in \eqref{eq:1}.
When correcting the prior prediction of the state variables $\bm{x}_c$ and $\bm{x}_a$ using the data $\bm{y}$, special consideration must be given to their cross-correlation structure, necessitating a careful design of the KF as outlined below. To simplify notation, we omit the dependency on the switching sequence, e.g., $A_c=A_c(\hat{s}_c(t))$).

First, we need to compute the prior state estimates of $\bm{x}_c$ and $\bm{x}_a$, denoted as $\hat{\bm{x}}_c^-$ and  $\hat{\bm{x}}_a^-$. The prior estimates are derived from the first two relations in \eqref{eq:1} as $\hat{x}^-_c(t)=A_c\hat{x}_c(t-1)$ and $\hat{x}^-_a(t)=A_a\hat{x}_a(t+1)$. With this regard, note that the switching sequence for each step has already been calculated. Successively, the measurement equation in \eqref{eq:1} allows us to perform posterior corrections $\hat{\bm{x}}_c$ and $\hat{\bm{x}}_a$ on the underlying prior estimates $\hat{\bm{x}}_c^-$ and  $\hat{\bm{x}}_a^-$ as follows:
\begin{align*}
	\hat{x}_c(t)&=\hat{x}_c^-(t)+K_c(y(t)-C_a\hat{x}_a^-(t)-C_c\hat{x}_c^-(t)),\\
	\hat{x}_a(t)&=\hat{x}_a^-(t)+K_a(y(t)-C_a\hat{x}_a^-(t)-C_c\hat{x}_c^-(t)),
\end{align*}
where $K_c\in\mathbb{R}^{n_{x_c}\times n_y}$ and $K_a\in\mathbb{R}^{n_{x_a}\times n_y}$ are the Kalman gains for the causal and non-causal states, respectively, whose design is critical for the effectiveness of the KF.
Before delving into the derivation of $K_c$ and $K_a$, let us first calculate the error covariance matrix for the prior state estimates based on the prior estimation errors $e_c^-(t)=x_c(t)-\hat{x}^-_c(t)$ and $e_a^-(t)=x_a(t)-\hat{x}^-_a(t)$, and the posterior estimation errors $e_c(t)=x_c(t)-\hat{x}_c(t)$ and $e_a(t)=x_a(t)-\hat{x}_a(t)$. Denote $P_c^-\triangleq\mathbb{E}[e_c^-(t)e_c^-(t)^\top],
P_a^-\triangleq\mathbb{E}[e_a^-(t)e_a^-(t)^\top],
P_c\triangleq\mathbb{E}[e_c(t)e_c(t)^\top],
P_a\triangleq\mathbb{E}[e_a(t)e_a(t)^\top].$ 
The Kalman gains are calculated to minimize the error covariance matrices of the posterior state estimates. The posterior estimation error can be rewritten as:
\begin{align*}
e_c(t)&=(I-K_cC_c)e_c^-(t)-K_cC_ae_a^-(t)-K_cv_m(t),\\
e_a(t)&=(I-K_aC_a)e_a^-(t)-K_aC_ce_c^-(t)-K_av_m(t),
\end{align*}
while the error covariance matrices of the state estimates:
\begin{align}
	P_c&=P_c^--P_c^-C_c^\top K_c^\top-K_cC_cP_c^-+K_cC_cP_c^-C_c^\top K_c^\top\nonumber\\
	&\hspace{2cm}+K_cC_aP_a^-C_a^\top K_c^\top+K_c\Sigma_mK_c^\top, \label{eq:1.5.1}\\
	P_a&=P_a^--P_a^-C_a^\top K_a^\top-K_aC_aP_a^-+K_aC_aP_a^-C_a^\top K_a^\top\nonumber\\
	&\hspace{2cm}+K_aC_cP_c^-C_c^\top K_a^\top+K_a\Sigma_mK_a^\top, \label{eq:1.5.2}
\end{align}
where the second equality in each derivation is established based on the independence of $e_c^-(t)$, $e_a^-(t)$ and $v_m(t)$. Note that minimizing the variances of $P_c$ and $P_a$ is equivalent to minimizing their traces. Therefore, given the unconstrained nature of such trace minimization, the optimal Kalman gains $K_c$ and $K_a$ can be found as:
\begin{align*}
	K_c&=(C_cP_c^-C_c^\top+C_aP_a^-C_a^\top+\Sigma_m)^{-1}(P_c^-C_c^\top),\\
	K_a&=(C_aP_a^-C_a^\top+C_cP_c^-C_c^\top+\Sigma_m)^{-1}(P_a^-C_a^\top).
\end{align*}

By substituting the Kalman gains above into \eqref{eq:1.5.1}--\eqref{eq:1.5.2}, the updated error covariance matrices can be obtained as: $P_c=(I-K_cC_c)P_c^-, P_a=(I-K_aC_a)P_a^-.$ By completing the steps of the modified KF, including the prediction, measurement update, and error covariance matrix update \cite{Angelo2010}, one can obtain all the posterior estimates of the state vectors $\bm{x}_c$ and $\bm{x}_a$, which are optimal state estimates based on the available measurements and prior knowledge. In addition, to ensure the convergence of the proposed state estimation method, we establish the following properties of the state estimates:

\begin{lemma}\label{lemma:boundendness}
	Let $\eta_c(t)=x_c(t)-A(\hat{s}_c(t))x_c(t-1)$, $\eta_a(t)=x_a(t)-A(\hat{s}_a(t))x_a(t+1)$, and $\eta_m(t)=y(t)-C_c(\hat{s}_c(t))x_c(t)-C_a(\hat{s}_a(t))x_a(t)$. There exist $\alpha_1$, $\alpha_2$, $\alpha_3 > 0$ so that $\|\eta_c(t)\|^2\le\alpha_1$, $\|\eta_a(t)\|^2\le\alpha_2$, and $\|\eta_m(t)\|^2\le\alpha_3$, for all $t\in\mathbb{Z}$.
\end{lemma}
The proof of the Lemma \ref{lemma:boundendness} is shown in \ref{sec:proofs}.

Lemma \ref{lemma:boundendness} states that the error of state estimation is bounded in the mean square sense, regardless of how the state trajectory evolves in time.
\begin{remark}
		In the E-step, the modified Kalman filter provides posterior state estimates along with their error covariance matrices. Under the Gaussian noise assumptions, the posterior distributions of states given the observations $\bm{y}$ are also Gaussian. Substituting these closed-form expectations (states, error covariance matrices) into \eqref{eq2.3}, the components in $Q(\theta,\theta^k)$ can be calculated.
	\end{remark}

\subsection{The M-step}
The second step in Algorithm~\ref{alg:EM} requires the maximization of $Q(\theta,\theta^k)$ to update the parameters estimate $\theta^k$:
\[
	\theta^{k+1}=\underset{\theta}{\arg\max}~Q(\theta,\theta^k).
\]
Let us first focus on the elements $\{\{\pi^c_i\}_{i=1}^{m_c},\{\pi^a_i\}_{i=1}^{m_a}\}$, and recall the objective function in \eqref{eq:1.4}. The $(k+1)$-th estimate of $\{\{\pi^c_i\}_{i=1}^{m_c},\{\pi^a_i\}_{i=1}^{m_a}\}$ can hence be obtained in closed-form by applying the first-order optimality conditions as follows:
\begin{align*}
	\pi^c_j=\frac{\sum_{t=1}^{T}w^c_{tj}}{\sum_{t=1}^{T}\sum_{j=1}^{m_c}w^c_{tj}},\quad	\pi^a_l=\frac{\sum_{t=1}^{T}w^a_{tl}}{\sum_{t=1}^{T}\sum_{l=1}^{m_a}w^a_{tl}}.
\end{align*}
Furthermore, the expression for the parameters $\{ \{\theta^c_i\}_{i=1}^{m_c},\{\theta^a_i\}_{i=1}^{m_a},\Sigma_m\}$ can be computed in closed-form by using the switching least-squares approach as follows:
\begin{align*}
	&A_c(j)=\underset{A_c(j)}{\arg\min}~\sum_{t=1}^{T}w^c_{tj}\Vert x_c(t)-\mu_2(t,j)\Vert^2,\\
	&A_a(l)=\underset{A_a(l)}{\arg\min}~\sum_{t=1}^{T}w^a_{tl}\Vert x_a(t)-\mu_3(t,l)\Vert^2,\\
	&(C_c(j),C_a(l))=\underset{(C_c(j),C_a(l))}{\arg\min}~\sum_{t=1}^{T}w^c_{tj}w^a_{tl}\Vert y(t)
	-\mu_1(t,j,l)\Vert^2.
\end{align*}
Then, the covariance matrices related to the disturbances $v_c$, $v_a$, and $v_m$ can also be estimated as:
{
	\begin{align*}
		&\Sigma_c(j)=\sum_{t=1}^{T} w^c_{tj}(x_c(t)-\mu_2(t,j))(x_c(t)-\mu_2(t,j))^\top,\\
		&\Sigma_a(l)=\sum_{t=1}^{T}w^a_{tl}(x_a(t)-\mu_3(t,l))(x_a(t)-\mu_3(t,l))^\top,\\
		&\Sigma_m=\sum_{t=1}^{T} w^c_{tj}w^a_{tl}(y(t)
		-\mu_1(t,j,l))(y(t)-\mu_1(t,j,l))^\top.
	\end{align*}
}
To show the convergence rate of the system matrices, we need the following definition of strong consistency of the parameter estimates. Recall that $\hat \theta$ is the estimate of $\theta$ made by exploiting $T$ samples. 
\begin{myDef}\textup{(\cite{Lai1982})} The estimate $\hat{\theta}$ is strongly consistent if 
	$
		\lim_{T\rightarrow\infty}\hat{\theta}=\theta.
	$
\end{myDef}

We are now ready to establish the convergence rate for $\hat{\theta}$. Due to the possible different active subsystems at time $t$, it is convenient to define the following partition of the considered time interval $\{1, \ldots, T\}$ as $\mathbb{T}^c_{j,T}=\{t\leq T\vert s_c(t)=j\}$ and $\mathbb{T}^a_{l,T}=\{t\leq T\vert s_a(t)=l\}$.
\begin{theorem}\label{th3}
	Under Standing Assumptions \ref{assumpt2} and \ref{assump1}. Let $W^c_{j,T}=\sum_{t\in\mathbb{T}^c_{j,T}} x_c(t)x_c^\top(t)$ and $W^a_{l,T}=\sum_{t\in\mathbb{T}^a_{l,T}}x_a(t)x_a^\top(t)$. Denote $\Delta^c_{j,T}=\mathcal{O}\left(\sqrt{\frac{\log(\lambda_{\textrm{max}}(W^a_{l,T}))}{\lambda_{\textrm{min}}(W^a_{l,T})}}\right)$, $\Delta^a_{l,T}=\mathcal{O}\left(\sqrt{\frac{\log(\lambda_{\textrm{max}}(W^a_{l,T}))}{\lambda_{\textrm{min}}(W^a_{l,T})}}\right)$, and $\Delta_T=\mathcal{O}\left(\frac{\log(T)}{T}\right)$. Then, the estimate $\hat{\theta}$ generated by Algorithm~\ref{alg:EM} is strongly consistent for any $\bm{s}_c\in\Lambda_c^T$ and $\bm{s}_a\in\Lambda_a^T$, and the convergence rates are
	\begin{align*}
		&\Vert \hat{A}_c(j)-A_c(j)\Vert_\infty\leq\Delta^c_{j,T},
		&\Vert \hat{A}_a(l)-A_a(l)\Vert_\infty\leq\Delta^a_{l,T},\\
		&\Vert \hat{C}_c(j)-C_c(j)\Vert_\infty\leq\Delta^c_{j,T},
		&\Vert \hat{C}_a(l)-C_a(l)\Vert_\infty\leq\Delta^a_{l,T}.\\
		&\Vert \hat{\Sigma}_c(j)-\Sigma_c(j)\Vert_\infty\leq\Delta_T,
		&\Vert \hat{\Sigma}_a(l)-\Sigma_a(l)\Vert_\infty\leq\Delta_T,\\
		&\Vert \hat{\Sigma}_m-\Sigma_m\Vert_\infty\leq\Delta_T.
	\end{align*}
\end{theorem}
The proof of Theorem \ref{th3} is shown in \ref{sec:proofs}.
\begin{remark}
	Theorem \ref{th3} gives data-dependent upper bounds for the estimation errors of the parameter matrices. In order to have a data-independent characterization of the convergence rate for adaptive control or reinforcement learning purposes, in the proof of Theorem~\ref{th3}, specifically equation \eqref{eq6.1}, we provide with the corresponding convergence rate of the parameter estimate $\hat{\theta}$, which is equal to $\mathcal{O}(\sqrt{\log(T)/T})$.
\end{remark}

\section{Numerical examples}\label{sec6}
We now verify the effectiveness of the proposed methodology on two simulation examples. In both cases, we note that the true switching sequences $\bm{s}_c$ and $\bm{s}_a$ are only used to verify the accuracy of the estimated switching sequences, i.e., $\hat{\bm{s}}_c$ and $\hat{\bm{s}}_a$. As performance index we make use of the mode match rate, defined as:
\[
	L_{\text{mr}}= \frac{1}{T}\sum_{t=1}^{T}\iota(s_c(t),\hat{s}_c(t)),
\]
where $\iota(\cdot,\cdot)$ denotes the standard indicator function, i.e., $\iota(s_c(t),\hat{s}_c(t))=1$ if $s_c(t)=\hat{s}_c(t)$, $0$ otherwise. 
\subsection{Example 1: Academic NCS-RSM}
For illustrative purposes, we start by considering a simple non-causal system described in \eqref{eq:1} with $m_c=m_a=2$ modes and main parameters reported in Table~\ref{tab:1} (refer to the ``True'' columns). The dimensions of the outputs, causal states, and non-causal states are $n_y=1,n_{x_c}=n_{x_a}=2$. The probabilities of all modes are $\pi^c_1=0.7,\pi^c_2=0.3,\pi^a_1=\pi^a_2=0.5$. The system is excited with white noise with zero mean and finite variance, and the data length is $T=10^4$.

\begin{figure}
	\centering
		\includegraphics[height=0.8\linewidth]{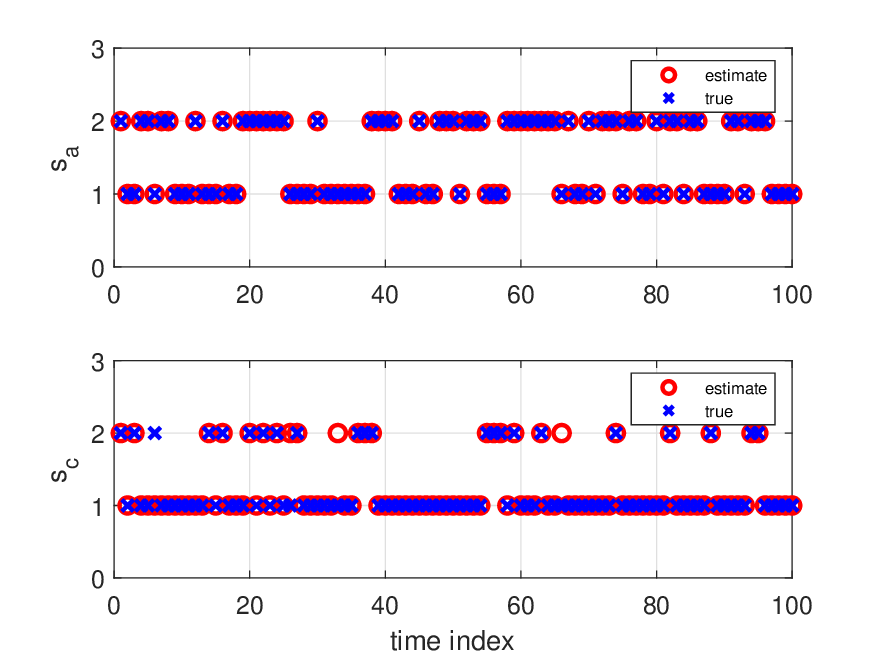} 
		\caption{The true (blue cross) and estimated (red circle) mode sequences over a certain time window of length $100$.}
		\label{fig:1}
\end{figure}

\begin{figure}
	\centering
	\includegraphics[height=0.8\linewidth]{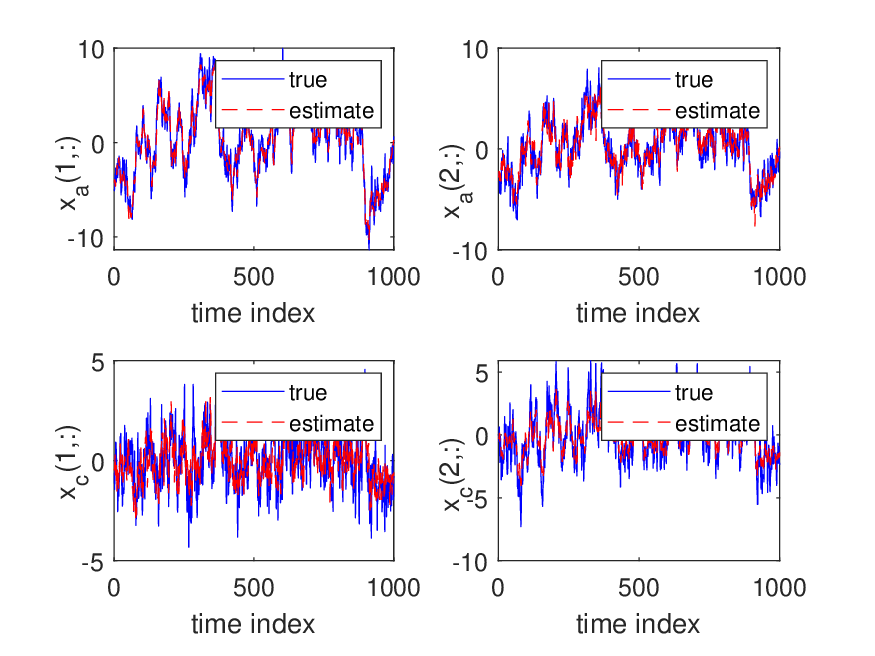} 
	\caption{Dynamical evolution of the true state variables $\bm{x}_c$ and $\bm{x}_a$ (solid blue line), and of the estimated ones $\hat{\bm x}_c$, and $\hat{\bm x}_a$ (dashed red lines).}
	\label{fig:2}
\end{figure}

The true and estimated parameters are reported in Table \ref{tab:1}, which clearly shows that the parameter estimates are very close to their true values. In Fig. \ref{fig:1} we report the partial estimation of the switching sequences $\bm{s}_c$ and $\bm{s}_a$, where the mode match rates are $97.4\%$ and $99.2\%$, respectively. Note that our method achieves an accurate parameter estimate, since each data point can be accurately assigned to the corresponding mode. 
To better validate the accuracy of the proposed algorithm in parameter estimation, Fig. \ref{fig:2} illustrates the estimated states using the modified KF. The relative estimation errors, defined as $\delta_c=\Vert x_c-\hat{x}_c\Vert^2/\Vert x_c\Vert^2$ ($\delta_a$ has the same structure), are $\delta_c=3.74\%$ and $\delta_a=3.14\%$, respectively.

\begin{table*}[!hbtp]
	\centering
	\renewcommand{\arraystretch}{1.2}
	\caption{The true and estimated system parameters}
	\label{tab:1}
	\begin{tabular}{|c|c|c|c|c|c|}\hline
		& True & Estimate &      & True & Estimate\\\hline
		$A_a(1)$&$\begin{bmatrix} 1 & 0 \\ 0 & 1 \end{bmatrix}$&    $\begin{bmatrix} 0.9681 & 0.0120 \\ 0.0142 & 0.9868 \end{bmatrix}$&$A_a(2)$&$\begin{bmatrix} 0.6 & 0.2 \\ 0.3 & 0.8 \end{bmatrix}$&$\begin{bmatrix} 0.6242 & 0.1992 \\ 0.3283 & 0.7738 \end{bmatrix}$\\\hline $A_c(1)$&$\begin{bmatrix} 1 & 0.2 \\ 0.3 & 0.8 \end{bmatrix}$&$\begin{bmatrix} 1.0131 & 0.2130 \\ 0.2849 & 0.8333 \end{bmatrix}$&$A_c(2)$&$\begin{bmatrix} 0.8 & 0.2 \\ 0.3 & 0.5 \end{bmatrix}$&$\begin{bmatrix} 0.8118 & 0.1899 \\ 0.3291 & 0.4784 \end{bmatrix}$\\\hline
		$C_a(1)$&$\begin{bmatrix} 0.2 & 0.6 \end{bmatrix}$&    $\begin{bmatrix} 0.2011 & 0.5962\end{bmatrix}$&$C_a(2)$&$\begin{bmatrix} 0.3 & 0.76 \end{bmatrix}$&$\begin{bmatrix} 0.2850 & 0.7677 \end{bmatrix}$\\\hline
		$C_c(1)$&$\begin{bmatrix} 0.3 & 0.7 \end{bmatrix}$&    $\begin{bmatrix} 0.2983 & 0.6979\end{bmatrix}$&$C_c(2)$&$\begin{bmatrix} 0.7 & 0.2 \end{bmatrix}$&$\begin{bmatrix} 0.7023 & 0.2029 \end{bmatrix}$\\\hline
		$\pi^c_1$&0.7&0.6963&$\pi^c_2$&0.3&0.3037\\\hline
		$\pi^a_1$&0.5&0.493&$\pi^a_2$&0.5&0.507\\\hline
		$\Sigma_a(1)$&$\begin{bmatrix} 1 & 0 \\ 0 & 1 \end{bmatrix}$&    $\begin{bmatrix} 1.1111 & -0.0711 \\ -0.0711 & 0.9865 \end{bmatrix}$&$\Sigma_a(2)$&$\begin{bmatrix} 1 & 0 \\ 0 & 1 \end{bmatrix}$&$\begin{bmatrix} 0.9307 & 0.0567 \\ 0.0567 & 1.0386 \end{bmatrix}$\\\hline
		$\Sigma_c(1)$&$\begin{bmatrix} 1 & 0 \\ 0 & 1 \end{bmatrix}$&    $\begin{bmatrix} 0.9773 & -0.0067 \\ -0.0067 & 0.9763 \end{bmatrix}$&$\Sigma_c(2)$&$\begin{bmatrix} 1 & 0 \\ 0 & 1 \end{bmatrix}$&$\begin{bmatrix} 1.0134 & -0.0001 \\ -0.0001 & 0.9850 \end{bmatrix}$\\\hline
		$\Sigma_m$&1&1.0049&&&\\\hline
	\end{tabular}
\end{table*} 

To validate the effectiveness of the proposed method, we compare it with the EM-based identification framework for a JMLS in \cite{Mark2022}, which was originally derived for causal switching systems. To accommodate the non-causal switching dynamics inherent in NCS-RSM, we extend the two-filter approach to a bidirectional filtering architecture consisting of a forward filter optimized for causal system components, and a backward filter specifically tailored for non-causal dynamics.
The length of the data is set to $T=10^4$. The transition matrix in \cite{Mark2022} is set to $\mathcal{T}=\left[\begin{smallmatrix}
	0.5&0.5\\
	0.5&0.5
\end{smallmatrix}\right]$, and the probability of the switching sequence in this paper is set to $\pi^c_1=\pi^c_2=\pi^a_1=\pi^a_2=0.5$. The subsystem match rates of the proposed method and \cite{Mark2022} are compared at different noise levels by assuming $\Sigma=\Sigma_c=\Sigma_a$. The identification accuracy of the switching sequences are shown in Table~\ref{tab:2}.
\begin{table}[!hbtp]
	\centering
	\caption{The mode match rates achieved by the EM algorithm in \cite{Mark2022} and by the proposed method.}
	\label{tab:2}
	\begin{tabular}{ccccc}\hline
		  & $L_{\text{mr}}(\bm{s})$ \cite{Mark2022} &  $L_{\text{mr}}(\bm{s}_c)$    & $L_{\text{mr}}(\bm{s}_a)$ \\\hline
		$\Sigma=0.00$& $100\%$&$100\%$ &$100\%$ &\\\hline 
		$\Sigma=0.01$&$99.5\%$ & $98.5\%$& $99.3\%$&\\\hline 
		$\Sigma=0.1$& $96.5\%$&$97.6\%$ &$99.1\%$ &\\\hline 
		$\Sigma=1$& $89.2\%$&$97.4\%$ &$99.2\%$ &\\\hline 
	\end{tabular}
\end{table} 

\begin{figure}
	\centering
	\includegraphics[height=0.8\linewidth]{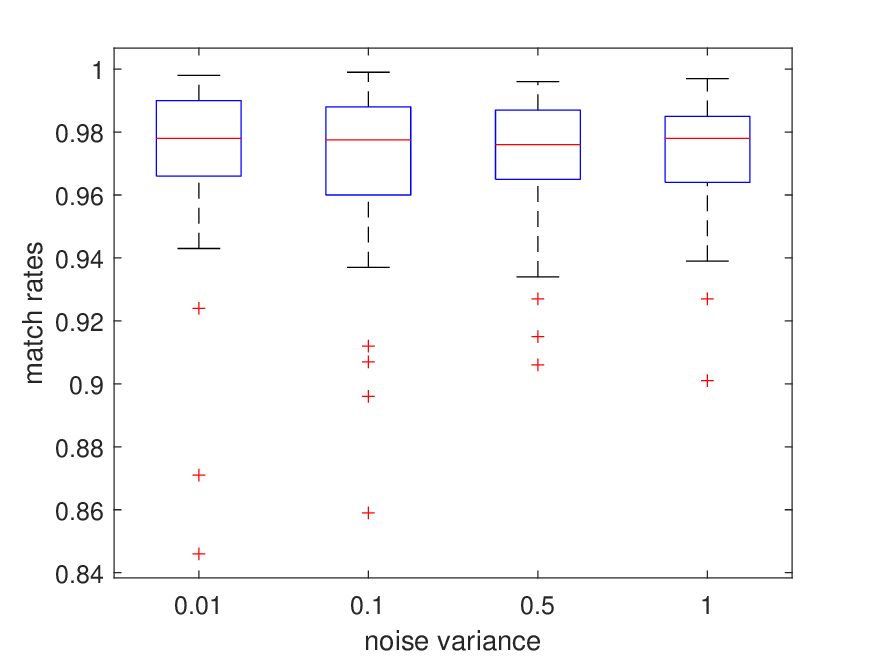} 
	\caption{Match rates obtained by the proposed algorithm for different noise levels.}
	\label{fig:3}
\end{figure}

To verify the robust performance of the proposed method against several noise levels, we run $100$ Monte Carlo experiments under four different noise conditions, i.e., $\Sigma\in\{0.01I,0.1I,0.5I,I\}$. In Fig. \ref{fig:3} we report the mean and the variance of the match rates in all the considered cases. We observe that the estimation accuracy of the switching sequence is not significantly affected by the noise variance, since even for high noise levels the estimation accuracy can still reach $98\%$ due to the excellent performance of the modified KF.

\subsection{Example 2: The Department Store Inventory Price Index }
In this subsection we adopt ``The Department Store Inventory Price Index"(DSIP) dataset from  The Bureau of Labor Statistics (BLS). These data come from inventory weighted price indices of goods carried by department stores.  

The dynamics of the DSIP are shaped by the interplay of both causal and non-causal factors. Traditional causal models focus on predicting future prices using historical data (e.g., supply-demand fluctuations, production costs), while the core value of the proposed NCS-ASM lies in historical data smoothing and dynamic interpretation. For instance, future expectations (e.g., pre-holiday inventory adjustments) influence historical price smooothing estimates through non-causal subsystems, correcting fluctuations caused by short-term market noise or measurement errors. Additionally, price dynamics often exhibit bidirectional feedback (e.g., interactions between current inventory and future restocking plans) and mode switching (e.g., seasonal patterns). Therefore, a NCS-RSM  model \eqref{eq:1} is suitable for describing the DSIP.

\begin{figure*}
	\centering
\begin{subfigure}{0.32\linewidth}
	\includegraphics[height=0.6\linewidth]{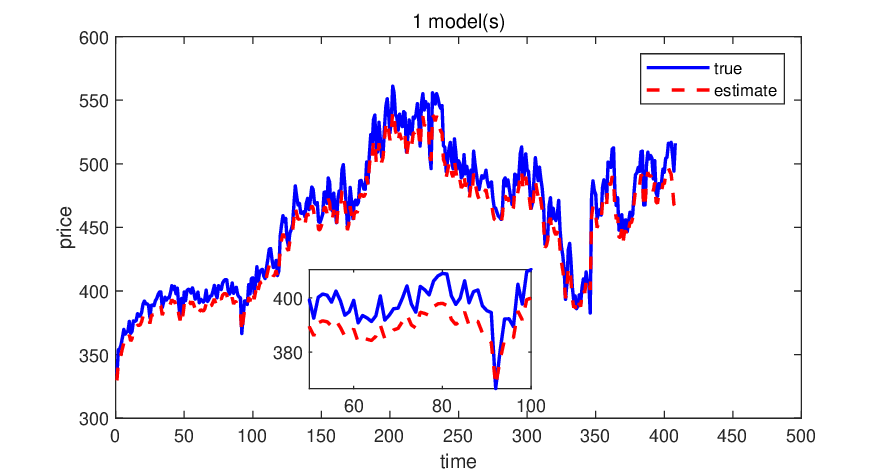} 
	\caption{}
\end{subfigure}
	\centering
\begin{subfigure}{0.32\linewidth}
		\includegraphics[height=0.6\linewidth]{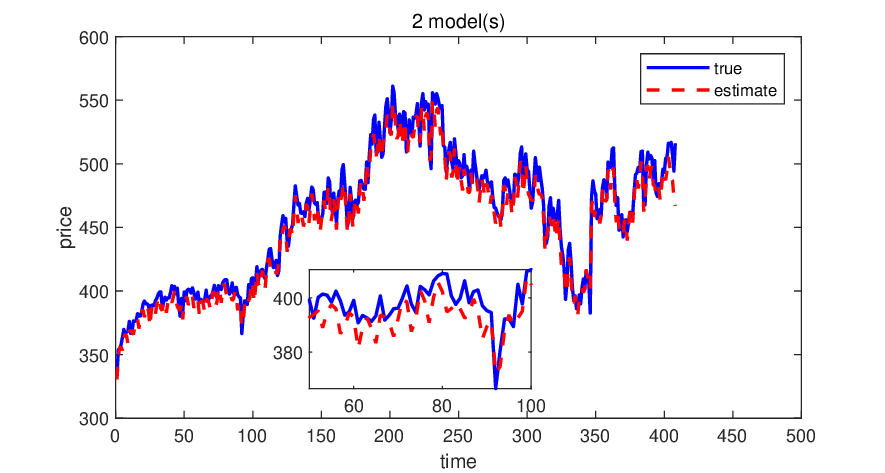} 
	\caption{}
\end{subfigure}
	\centering
\begin{subfigure}{0.32\linewidth}
	\includegraphics[height=0.6\linewidth]{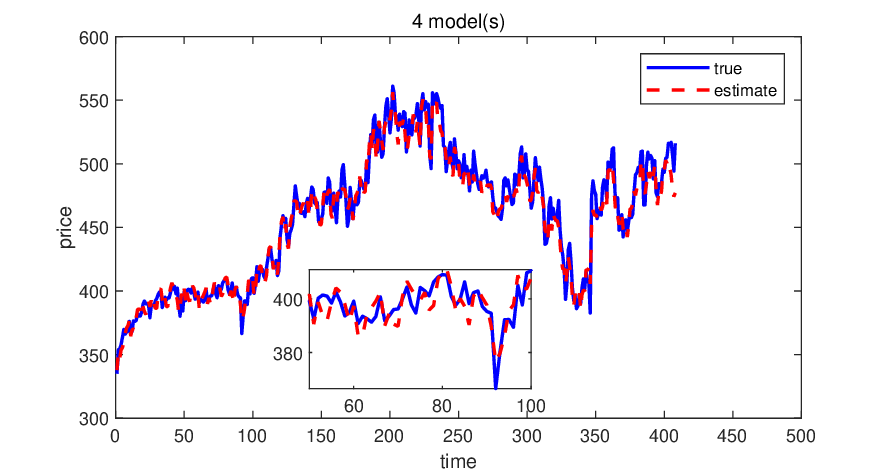} 
	\caption{}
\end{subfigure}
	\centering
	\caption{The smoothing estimated prices and the true prices with different numbers of the subsystem. (a) $\bm{s}_c=\bm{s}_a$ and $m_c=m_a=1$; (b) $\bm{s}_c\neq \bm{s}_a$ and $m_c=m_a=1$; (c) $\bm{s}_c\neq \bm{s}_a$ and $m_c=m_a=2$}\label{fig:5}
\end{figure*}

In Fig. \ref{fig:5} we show the true prices and the smoothing estimated prices with different number of subsystems. The smoothing estimation errors $\delta=\Vert \bm{y}-\hat{\bm{y}}\Vert/\Vert \bm{y}\Vert$ with different number of subsystems are shown in Table \ref{tab:3}. Specifically, we can infer that the larger the number of subsystems, the more accuracy the smoothing estimation becomes.
\begin{table}[!hbtp]
	\centering
	\caption{The smoothing estimation errors against different number of subsystems.}
	\label{tab:3}
	\begin{tabular}{cccc}\hline
		switching sequence & $\#$ of $\bm{s}_c$ &  $\#$ of $\bm{s}_a$& $\delta$  \\\hline
		$\bm{s}_c=\bm{s}_a$& $m_c=1$&$m_a=1$&0.0249 \\\hline 
		$\bm{s}_c\neq \bm{s}_a$& $m_c=1$&$m_a=1$&0.0195 \\\hline
		$\bm{s}_c\neq \bm{s}_a$& $m_c=2$&$m_a=2$& 0.0188\\\hline
	\end{tabular}
\end{table} 

In conclusion, from Fig. \ref{fig:5} and Table \ref{tab:3} we note that switching systems with a larger number of modes can obtain the more accurate smoothing estimate of inventory levels, which can offer a guide restocking decisions.

\section{Conclusion}\label{sec8}
We have proposed an expectation-maximization framework for identifying non-causal systems with random switching modes. In the E-step, we have embedded the reconstructed switching sequence into the modified Kalman filter so that the proposed algorithm can handle the joint state variable estimation for the causal and non-causal parts. Furthermore, in M-step we have developed a switching least-squares algorithm  that can get the parameter estimates in closed-form. From a technical perspective, we have established the convergence of our identification methodology, also deriving an upper bound $\mathcal{O}(\sqrt{\log(T)/T})$ for the parameter errors. 

Note that the identification algorithm proposed in this paper can be adapted to the identification of switching linear descriptor systems with minor modifications, since a descriptor state-space model can be represented in the mixed causal and non-causal form. When the subsystems are nonlinear, however, the identification task becomes more challenging, thus posing greater difficulties. This aspect will be further investigated in our future work. In addition, addressing the joint identification of structured subsystems and piecewise constant switching sequences is an interesting future research direction.



\begin{thebibliography}{99}     

\bibitem{Tan2023} K. Tan, W. J. Parquette, \& M. Tao. (2023). A predictive algorithm for maximum power point tracking in solar photovoltaic systems through load management. Solar Energy, 265, 112127.

\bibitem{Liao2024} S. Liao, Y. Wu, K. Ma, \& Y. Niu, (2024). Ant Colony Optimization With Look-Ahead Mechanism for Dynamic Traffic Signal Control of IoV Systems. IEEE Internet of Things Journal, 11(1), 366-377.
\bibitem{Carloni2007} R. Carloni, R. G. Sanfelice, A. R. Teel, \& C. Melchiorri. (2007). A hybrid control strategy for robust contact detection and force regulation. In Proc. American Control Conf., New York City, USA, 1461-1466.
\bibitem{Liu2016} C. Liu, J. Li, J. Wang, \& Y. Tian. (2016). Supply Chain Simulation with Switching Adaptive Model Predictive Control Methodology. Proceedings of the 2016 International Conference on Sensor Network and Computer Engineering (pp. 639-646).

\bibitem{El2023} Andrews, B., Calder, M., \& Davis, R.A. (2007). Maximum likelihood estimation for $\alpha$-stable autoregressive processes. Ann. Statist. 37, 1946–1982.
\bibitem{Schlegl2003} T. Schlegl, M. Buss, \& G. Schmidt. (2003). A hybrid systems approach toward modeling and dynamical simulation of dextrous manipulation. IEEE/ASME Trans. on Mechatronics, 8(3), 352-361.




\bibitem{Chan2008} Chan, A. B., \& Vasconcelos, N. (2008). Modeling, clustering, and segmenting video
with mixtures of dynamic textures. IEEE Transactions on Pattern Analysis and
Machine Intelligence, 30(5), 909–926.

\bibitem{Ferrari2003} Ferrari-Trecate, G., Muselli, M., Liberati, D., \& Morari, M. (2003). A clustering
technique for the identification of piecewise affine systems. Automatica, 39(2),
205–217.

\bibitem{Garulli2012} Garulli, Andrea, Paoletti, Simone, \& Vicino, Antonio (2012). A survey on switched
and piecewise affine system identification. In 16th IFAC symposium on system
identification, Brussels, Belgium (pp. 344–355).

\bibitem{Bianchi2021} Bianchi, Federico, Breschi, Valentina, Piga, Dario, \& Piroddi, Luigi (2021). Model
structure selection for switched NARX system identification: A randomized
approach. Automatica, 125, Article 109415.

\bibitem{Mark2022} Mark P. Balenzuela, Adrian G. Wills, Christopher Renton, \& Brett Ninness. (2022). Parameter estimation for Jump Markov Linear Systems. Automatica, 135 109949.

\bibitem{Bemporad2018} Alberto Bemporad, Valentina Breschi, Dario Piga, \& Stephen P. Boyd. (2018). Fitting jump models. Automatica, 96, 11-21.

\bibitem{Piga2020} Dario Piga, Valentina Breschi, \& Alberto Bemporad. (2020). Estimation of jump Box–Jenkins models. Automatica, 120 109126.

\bibitem{Sayedana2024} Borna Sayedana, Mohammad Afshari, Peter E. Caines, \& Aditya Mahajan. (2024). Strong Consistency and Rate of Convergence of Switched Least Squares System Identification for Autonomous Markov Jump Linear Systems. IEEE transactions on Automatic Control, 1-8.

\bibitem{Anna2018} Anna Scampicchio, Alberto Giaretta, \& Gianluigi Pillonetto. (2018). Nonlinear Hybrid Systems Identification using Kernel-Based Techniques. In IFAC-PapersOnline, 51(15), 269-274.

\bibitem{Angelo2010} Angelo Alessandri, Marco Baglietto, \& Giorgio Battistelli. (2010). A maximum-likelihood Kalman filter for switching discrete-time linear systems. Automatica, 46, 1870-1876.

\bibitem{Cao2016} X. Cao, Bruce Stephen, Ibrahim F. Abdulhadi, C. D. Booth, \& G. M. Burt. (2016). Switching Markov Gaussian Models for Dynamic Power System Inertia Estimation. IEEE Transactions on Power Systems, 31(5), 3394-3403.



\bibitem{Yang2012} Miin-Shen Yang, Chien-Yo Lai, \& Chih-Ying Lin. (2012), A robust EM clustering algorithm for Gaussian mixture models. Pattern Recognition, 45(11), 3950-3961.



\bibitem{Verhaegen1996} Verhaegen, M. (1996). A subspace model identification solution to the identification of mixed causal, anti-causal LTI systems. SIAM Journal on Matrix Analysis and Applications, 17(2), 332–347.

\bibitem{Fang2024} X. Fang, \& T. Chen. (2024). On kernel design for regularized non-causal system identification. Automatica, 159, 111335.

\bibitem{Blanken2020} Blanken, L., \& Oomen, T. (2020). Kernel-based identification of non-causal systems with application to inverse model control. Automatica, 114.

\bibitem{sta_as1} T. E. Duncan, \& B. Pasik-Duncan. (1990). Adaptive control of continuoustime linear stochastic systems. Math. Control signals systems, 3(1), 45–60.
\bibitem{sta_as2} M. K. S. Faradonbeh, A. Tewari, \& G. Michailidis. (2020). On adaptive linear–quadratic regulators. Automatica, 117, 108982.

\bibitem{Long2023} Jiamei Long, Yuqian Guo, \& Weihua Gui. (2023). Mean square stability of discrete-time linear systems with random impulsive disturbances. Science China Information Sciences, 66(6), 169203.
\bibitem{Cong2024} Shen Cong. (2024). Almost sure stability criteria for linear Markovian switching systems. ISA Transactions, 146, 285-290.

\bibitem{Dempster1977} Dempster, Arthur P., Laird, Nan M., \& Rubin, Donald B. (1977). Maximum likelihood from incomplete data via the EM algorithm. Journal of the Royal Statistical Society. Series B. Statistical Methodology, 1–38.
\bibitem{Andrea2020} Andrea Ruggieri, Francesco Stranieri, Fabio Stella, \& Marco Scutari. (2020). Hard and Soft EM in Bayesian Network Learning from Incomplete Data. Algorithms, 13(12), 329.
\bibitem{Pal2024} Pal, S., \& Heumann, C. (2024). Gaussian mixture model with modified hard EM algorithm in clustering problems. In Statistical Modeling and Applications on Real-Time Problems (pp. 153-179). CRC Press.

%
%
%
%
%
%

\bibitem{Lai1982} T. L. Lai, \& C. Z. Wei. (1982). Least squares estimates in stochastic regression
models with applications to identification and control of dynamic
systems. Ann. Statist., 10(1), 154–166.

\bibitem{PE2018} P. E. Caines. (2018). Linear stochastic systems. SIAM.
\bibitem{Chen1986} H. F. Chen, \& L. Guo. (1986). Convergence rate of least-squares identification and adaptive control for stochastic systems. Int J Control, 44(5), 1459–1476.





\end{thebibliography}

\appendix
\section{Technical proofs}\label{sec:proofs}
\textit{Proof of Lemma~\ref{lemma:1}:} The log likelihood difference between the $\theta$ and $\theta^{k}$ can be expressed as 
\begin{align*}
	\ln \mathbb P_\theta(\bm{y})-\ln \mathbb P_{\theta^k}[\bm{y}]=Q(\theta,\theta^k)-&Q(\theta^k,\theta^k)\\
	&+V(\theta,\theta^k)-V(\theta^k,\theta^k),
\end{align*}
where the difference $V(\theta,\theta^k)-V(\theta^k,\theta^k)$ coincides with the Kullback–Leibler distance that possess an important property, i.e., being non-negative. Therefore, the maximization of $Q(\theta,\theta^k)$ can yield an increase in the log-likelihood function $\ln \mathbb P_\theta(\bm{y})$, namely
\[
	Q(\theta,\theta^{k+1})\geq Q(\theta,\theta^{k})\Rightarrow \ln \mathbb P_{\theta^{k+1}}[\bm{y}]\geq\ln \mathbb P_{\theta^k}[\bm{y}],
\]
thus concluding the proof.
\qedsymbol

\emph{Proof of Lemma~\ref{lemma:boundendness}:} Only the boundedness of $\eta_c(t)$ will be proven in detail, since that of $\eta_a(t)$ and $\eta_m(t)$ can be derived in a similar way.

First, we note that $x_c(t-1)$ can be equivalently expressed  as follows:
\[
	x_c(t-1)=\varphi_1(\bm{s}_c)x_c(1)+\varphi_2(\bm{s}_c)\bm{v}_c(1:t-1),
\]
where $\bm{v}_c(1:t-1)\triangleq[v_c(1),\cdots,v_c(t-1)]$, $\varphi_1(\bm{s}_c)$ and $\varphi_2(\bm{s}_c)$ are shown as follows:
\begin{align}
	\varphi_1(\bm{s}_c)&=&A_c(s_c(2))+A_c(s_c(3))A_c(s_c(2))+\cdots+A_c(s_c(t-1))A_c(s_{t-2})\cdots A_c(s_c(2))\\
	\varphi_2(\bm{s}_c)&=&\begin{bmatrix}
		1+A_c(s_c(2))+A_c(s_c(3))A_c(s_c(2))+\cdots+A_c(s_c(t-1))\cdots A_c(s_c(2))\\
		1+A_C(s_c(3))+\cdots+A_c(s_c(t-1))\cdots A_c(s_c(3))\\
		\vdots\\
		1+A_c(s_c(t-1))\\
		1
	\end{bmatrix}^\top
\end{align}
Both matrices are uniquely determined by the switching sequence $\bm{s}_c$ and system matrices $A_c$. Then, one obtains that:

\begin{align*}
	\eta_c(t)&=x_c(t)-A(\hat{s}_c(t))x_c(t-1)\\
	&=(A_c(s_c(t))-A(\hat{s}_c(t))) x_c(t-1)+v_c(t)\\
	&=(A_c(s_c(t))-A(\hat{s}_c(t)))\varphi_1(\bm{s}_c)x_c(1)\\
	&\hspace{3.6cm}+\varphi_3(\bm{s}_c)\bm{v}_c(1:t),
\end{align*}
where $\varphi_3(\bm{s}_c)=[\varphi_2(\bm{s}_c),1]$. Passing to the (squared) norm in the expression above we note that, in view of the fact that the noise $\bm{v}_c$ has a bounded covariance, the last term is bounded too. For what concerns the first term, instead, we have:
\[
	\Vert [A_c(s_c(t))-A(\hat{s}_c(t))]\varphi_1(\bm{s})x_c(1)\Vert^2\leq\lambda_1\Vert x_c(1)\Vert^2,
\]
where 
\begin{align*}
	\lambda_1\triangleq\lambda_{\textrm{max}}(\varphi^\top_1(\bm{s})&(A_c(s_c(t))-A(\hat{s}_c(t)))^\top\\
	&\hspace{1.5cm}(A_c(s_c(t))-A(\hat{s}_c(t)))\varphi_1(\bm{s})),
\end{align*}
which concludes the proof.
\qedsymbol

\emph{Proof of Theorem~\ref{th3}:} In the interest of space, we establish the convergence rate for $\hat{A}_a(l)$ only, since the other bounds on the system matrices can be derived similarly. We start by introducing two necessary lemmas. In particular, the following result holds true by virtue of Standing Assumptions \ref{assumpt2} and \ref{assump1}:
\begin{lemma}\textup{(\cite[Lemma~3]{Sayedana2024})}\label{lemma3}
	The following asymptotic relations hold true almost surely (a.s.):
	\begin{align*}
		&\left\lVert\sum_{i=1}^{T}A(s_c(i))x_c(i)v_c^\top(i)+v_c(i)x_c^\top(i)A(s_c(i))\right\rVert=o(T),\\
		&\left\lVert \sum_{i=1}^{T}A(s_a(i))x_a(i)v_a^\top(i)+v_a(i)x_a^\top(i)A(s_a(i))\right\rVert=o(T).
	\end{align*}
\end{lemma}
The proof extends \cite[Lemma 3]{Sayedana2024} by treating the non-causal dynamics \eqref{eq:1} as a time-reversed causal process. By Standing Assumption 1, the reversed process preserves stability in the average sense. The martingale property of $v_a(t)$ (Standing Assumption 2) ensures the applicability of the covariance analysis in \cite{Sayedana2024}. 

Next, we report a lemma whose validity follows from Standing Assumption~\ref{assump1}:
\begin{lemma}\textup{(\cite{Lai1982})}\label{lemma1}
	The standard least-squares solution can be expressed as
	$
	\hat{A}_a(l)=\arg\min_{A_a(l)}\Vert x_a(t)-A_a(l) x_a(t+1)\Vert^2, \quad t\in\mathbb{T}^a_{l,T},
	$
	for all $l=1,\cdots,m_a$. If
	\begin{enumerate}
		\item[(C1)] $\lambda_{\textrm{min}}(W^a_{l,T})\rightarrow\infty$ a.s., and
		\item[(C2)] $\log\lambda_{\textrm{max}}(W^a_{l,T})=o(\lambda_{\textrm{min}}(W^a_{l,T}))$ a.s.,
	\end{enumerate}
	then the least-squares estimate $\hat{A}_a(l)$ is strongly consistent with convergence rate
	\[
	\Vert \hat{A}_a(l)-A_a(l)\Vert_\infty=\mathcal{O}\left(\sqrt{\frac{\log(\lambda_{\textrm{max}}(W^a_{l,T}))}{\lambda_{\textrm{min}}(W^a_{l,T})}}\right)~a. s.
	\]
\end{lemma}

In view of Lemma~\ref{lemma1}, sufficient conditions for establishing the convergence rate of $\hat{A}_a(l)$ are (C1) $\lambda_{\textrm{min}}(W^a_{l,T})\rightarrow\infty$, a.s., and (C2) $\log\lambda_{\textrm{max}}(W^a_{l,T})=o(\lambda_{\textrm{min}}(W^a_{l,N}))$, a.s.. We therefore have to show that these two conditions are verified in our case. Then, for what concerns (C1), one has:
\begin{align*}
	x_a(t)x_a(t)^\top&=(\hat{A}_a(l)x_a(t+1)+v_a(t))\\
	&\hspace{2.7cm}(\hat{A}_a(l)x_a(t+1)+v_a(t))^\top\\
	&=\hat{A}_a(l)x_a(t+1)x_a^\top(t+1)\hat{A}^\top_a(l)\\
	&\hspace{1cm}+2v_a(t)x^\top_a(t+1)\hat{A}^\top_a(l)+v_a(t)v_a^\top(t).
\end{align*}
Since $\hat{A}_a(l)x_a(t+1)x_a^\top(t+1)\hat{A}^\top_a(l)$ is a positive semidefinite matrix, by relying on Lemma \ref{lemma3} we can infer that
\begin{align*}
	W^a_{l,T}&=\sum_{t\in\mathbb{T}^a_{l,T}}x_a(t)x^\top_a(t)\\
	&\succcurlyeq\sum_{t\in\mathbb{T}^a_{l,T}}v_a(t)v_a^\top(t)+x_a(T)x^\top_a(T)\\
&+\sum_{t\in\mathbb{T}^a_{l,T}}(\hat{A}_a(l)x_a(t\!+\!1)v^\top_a(t)\!+\!v_a(t)x^\top_a(t\!+\!1)\hat{A}^\top_a(l))\\
	&\succcurlyeq\sum_{t\in\mathbb{T}^a_{l,T}}v_a(t)v_a^\top(t)+o(T).
\end{align*}
Then, we readily obtain:
\begin{align*}
	\lim_{\vert \mathbb{T}^a_{l,T}\vert\rightarrow\infty}\inf&\frac{\sum_{t\in\mathbb{T}^a_{l,T}}x_a(t)x^\top_a(t)}{\vert\mathbb{T}^a_{l,T}\vert}\\
	&\succcurlyeq\lim_{\vert \mathbb{T}^a_{l,T}\vert\rightarrow\infty}\inf\frac{\sum_{t\in\mathbb{T}^a_{l,T}}v_a(t)v_a^\top(t)}{\vert\mathbb{T}^a_{l,T}\vert}\succ0.
\end{align*}
Therefore, we can conclude that $\lambda_{\textrm{min}}(W^a_{l,T})\rightarrow\infty$ a.s..

To prove (C2) we note that:
\begin{align*}
	\lambda_{\textrm{max}}(\textstyle\sum_{t\in\mathbb{T}^a_{l,T}}x_a(t)x^\top_a(t))&\leq \textrm{tr}(\textstyle\sum_{t\in\mathbb{T}^a_{l,T}}x_a(t)x^\top_a(t))\\
	&\leq \sum_{i=1}^{T}\Vert x_a(t)\Vert^2=\mathcal{O}(N),
\end{align*} 
where the last equality follows in view of the stability, in average sense, of the NCS-RSM in \eqref{eq:1}.
Then, one can readily obtain that
\begin{eqnarray}\label{eq6.1}
	\lim_{T\rightarrow\infty}\frac{\log(\lambda_{\textrm{max}}(W^a_{l,T}))}{\lambda_{\textrm{min}}(W^a_{l,T})}\leq\lim_{T\rightarrow\infty}\frac{\log(T)}{\vert\mathbb{T}_{i,T}\vert}=\frac{\log(T)}{\mathcal{O}(T)}=0.\nonumber\\
\end{eqnarray}

We are now able to establish the convergence rate for the covariance matrices. Specifically, we will give the detailed proof for $\hat{\Sigma}_c(j)$ only, since the remaining ones follow similarly.

From the NCA-ASM in \eqref{eq:1}, the true covariance matrix for $v_c$ can be expressed as:
\begin{align*}
	\Sigma_c(j)=\frac{1}{\vert\mathbb{T}^c_{j,T}\vert}\sum_{t\in\mathbb{T}^c_{j,T}}(x_c(t)-&A_c(j)x_c(t-1))\\&(x_c(t)-A_c(j)x_c(t-1))^\top.
\end{align*}
Then, the estimation error can take the following form: 
\begin{align*}
	\hat{\Sigma}_c(j)-\Sigma_c(j)=&\frac{1}{\vert\mathbb{T}^c_{j,T}\vert}\sum_{t\in\mathbb{T}^c_{j,T}}((A_c(j)-\hat{A}_c(j))x_c(t-1))\\
	&\hspace{2cm}((A_c(j)-\hat{A}_c(j))x_c(t-1))^\top.
\end{align*}
Therefore, the convergence rate for $\hat{\Sigma}_c(j)$ reads as:
\begin{align*}
	\Vert\hat{\Sigma}_c(j)-\Sigma_c(j)\Vert_\infty\leq& \frac{\sum_{t\in\mathbb{T}^c_{j,T}}x_c(t-1)x_c(t-1)^\top}{\vert\mathbb{T}^c_{j,T}\vert}\\
	&\Vert (A_c(j)-\hat{A}_c(j))(A_c(j)-\hat{A}_c(j))^\top\Vert_\infty\\
	&\leq\mathcal{O}\left(\frac{\log(T)}{T}\right),
\end{align*}
which completes the proof.
\qedsymbol

\end{document}